# Two-Dimensional Node-Line Semimetals in a Honeycomb-Kagome Lattice

J. L. Lu[1], W. Luo[2,3], X. Y. Li[2], S. Q. Yang[2], J. X. Cao[1], X. G. Gong[2,3], H. J. Xiang[2,3]*

[1]*Department of Physics, Xiangtan University, Xiangtan, Hunan, 411105, P.R. China*

[2]*Key Laboratory of Computational Physical Sciences (Ministry of Education), State Key Laboratory of Surface Physics, and Department of Physics, Fudan University, Shanghai 200433, P. R. China*

[3]*Collaborative Innovation Center of Advanced Microstructures, Nanjing 210093, P. R. China*

*email: hxiang@fudan.edu.cn

**Abstract**

Recently, the concept of topological insulators has been generalized to topological semimetals, including three-dimensional (3D) Weyl semimetals, 3D Dirac semimetals, and 3D node-line semimetals. In particular, several compounds (e.g., certain three-dimensional graphene networks, $Cu_3PdN$, $Ca_3P_2$) were discovered to be 3D node-line semimetals, in which the conduction and the valence bands cross at closed lines in the Brillouin zone. Except for the two-dimensional (2D) Dirac semimetal (e.g., in graphene), 2D topological semimetals are much less investigated. Here, we propose the new concept of a 2D node-line semimetal and suggest that this state could be realized in a new mixed lattice (we name it as HK lattice) composed by kagome and honeycomb lattices. We find that $A_3B_2$ (A is a group-IIB cation and B is a group-VA anion) compounds (such as $Hg_3As_2$) with the HK lattice are 2D node-line semimetals due to the band inversion between cation s orbital and anion $p_z$ orbital. In the presence of buckling or spin-orbit coupling, the 2D node-line semimetal state may turn into 2D Dirac semimetal state or 2D topological crystalline insulating state.

PACS numbers: 73.20.At, 71.55.Ak, 71.20.−b

Due to the unique band structure, topological insulator (TI) has drawn broad attention in recent years [1-4]. It has a bulk energy gap between the valence and conduction bands which is similar to ordinary insulators, but contains a nontrivial gapless boundary state that is inspired by its bulk topological states. Similar to TIs, a new topological state of metal has been proposed, marked as topological semimetal state. In topological semimetals, the valence and conduction bands cross near the Fermi level to form band touch points and exhibit new topological quantum states. Up to now, according to the property of band touch points, three-dimensional (3D) topological semimetals have been divided into three kinds, i.e., Dirac semimetal (DSM) and Weyl semimetal (WSM) that possess discrete band touch points, and node-line semimetal (NLS) in which the band touch points form a closed ring in momentum space. Wan *et al.* first proposed that pyrochlore iridates (such as $Y_2Ir_2O_7$) are magnetic WSM where the time-reversal symmetry is broken [5]. Subsequently, Xu *et al.* has proposed that ferromagnetic $HgCr_2Se_4$ possesses a single pair of Weyl fermions with opposite chirality separated in momentum space [6]. Spatial inversion broken WSM state was also discovered in nonmagnetic materials including TaAs, TaP, NbAs and NbP [2, 7-11]. When two opposite chiral Weyl fermions meet, the DSM state with a four-fold degenerate Dirac node protected by both inversion and time-reversal symmetries may appear. $Na_3Bi$ [12, 13] and $Cd_3As_2$ [14-18] have been predicted theoretically and verified experimentally to be 3D DSM semimetal.

In 2011, Burkov *et al.* proposed the concept of 3D NLS with broken time-reversal symmetry and gave an explicit model realization of 3D NLS in a normal insulator-TI superlattice structure [19]. The 3D NLS state was recently extended to the case of time-reversal invariant systems [20, 21]. Three-dimensional carbon allotrope materials with a negligible SOC effect such as Mackay-Terrones crystals [20], hyperhoneycomb lattices [22], and the interpenetrated graphene network [23] were proposed to be time-reversal invariant 3D NLS. In addition, the cubic antiperovskite material $Cu_3PdN$ [24, 25], $Ca_3P_2$ [26, 27], non-centrosymmetric superconductor $PbTaSe_2$ [28] and $TlTaSe_2$ [29] were also predicted to display such exotic state. There are drumhead-like surface flat bands on the surface of 3D NLSs. When SOC is taken into account, each node-line ring may drive into a pair of Dirac nodes or Weyl nodes [24, 25].

In contrast to the case of 3D topological semimetals for which a comprehensive understanding is now achieved, the research field of two-dimensional topological semimetals is still under development. When the SOC effect is neglected, graphene can be seen as a 2D Dirac semimetal [30]. Young and Kane proposed the existence of three possible distinct 2D Dirac semimetal phases by using a two-site tight binding model [31]. A Lieb lattice with intra-unit-cell and suitable nearest-neighbor hopping terms between the spinless fermions was proposed to be a 2D Chern semimetal with a single Dirac-like point [32]. However, the concept of the 2D NLS has never been proposed and the corresponding material has not been discovered.

In the present work, based on effective model analysis and first-principles calculations, we put forward a new concept of the 2D NLS and show that this novel state can be realized in a new composite lattice (hereafter referred to as the HK lattice) with interpenetrating kagome and honeycomb lattices. Furthermore, first-principles calculations show that a series of compounds (e.g, $Hg_3As_2$) with the HK lattice are indeed 2D NLSs in which the band inversion happens between Hg-s orbital and As-$p_z$ orbital. When considering SOC, a tiny band gap at the node-line ring can be induced, resulting in a novel topological crystalline insulator without parity inversion.

**2D NLS in the HK lattice.** It is well-known that the band crossing in a 2D system could happen at a special point in the 2D Brillouin zone. For example, the $\pi$ and $\pi^*$ bands in graphene cross each other to form Dirac points at high symmetry points $K$ and $K'$. Here, we will address the interesting question whether the touching points can form a node-ring in a 2D system. For simplicity, we consider the spinless case (without the SOC effect). If two bands happen to cross each other at a k-point, there are two possibilities: Either these two bands belong to two different one-dimensional irreducible representations, or they together form the basis of a 2D irreducible representation. Among the k-points of a node-ring in the 2D Brillouin zone, some of them must be general k-points with the lowest point group symmetry. In the

3D case, a general k-point can only have the $C_1$ point group symmetry. In contrast, a general k-point in the 2D case could have the Cs point group symmetry if the 2D system have an in-plane mirror (or glide mirror) symmetry. Since the Cs point group only has two one-dimensional irreducible representations ($A'$ and $A''$), the two bands in a 2D NLS have to belong to $A'$ and $A''$ representations respectively. Therefore, to realize a 2D NLS, the material should have an in-plane mirror (or glide mirror) symmetry and the two bands near the Fermi level should transform differently under the mirror symmetry.

We find that the 2D NLS state may be present in a composite lattice (i.e, HK lattice) with interpenetrating kagome and honeycomb lattices. Let's start from the simple tight-binding model of a honeycomb lattice. In Figure 1a, we show the top view of a planar honeycomb lattice with the $D_{6h}$ point group. There are two sites per primitive cell and each site is associated with a $p_z$ orbital. For simplicity, we only consider the nearest neighboring (NN) interaction. Similar to the graphene case, we can obtain the eigenvalues of two bands with

$$E_{1,2} = E_p \pm t_p \sqrt{3 + 2\cos(2k_x) + 2\cos(k_x + \sqrt{3}k_y) + 2\cos(\sqrt{3}k_y - k_x)}$$

Here, $E_p$ and $t_p$ is the on-site energy of the $p_z$ orbital and the hopping between NN $p_z$ orbitals, respectively. It is well-known that there are Dirac points ($K$ and $K'$) in the band structure [30]. Another interesting observation from the expression of the eigenvalues is that the upper (lower) anti-bonding (bonding) band has the maximum (minimum) at $\Gamma$. For the kagome lattice, there are three sites in the primitive cell. Each site has one s orbital and only the NN interaction is taken into consideration. The eigenvalues of the three bands are

$$E_1 = E_s - 2t_s,$$

$$E_{2,3} = \mp t_s \sqrt{3 + 2\cos(2k_x) + 2\cos(k_x + \sqrt{3}k_y) + 2\cos(\sqrt{3}k_y - k_x)} + t_s + E_s,$$

where $E_s$ and $t_s$ are the on-site energy of the s orbital and the NN hopping parameter (negative for the s orbital), respectively. The topmost band (also see Fig. 1e) is flat, which is characteristic for the Kagome lattice [33, 34]. The other two lower bands have the same dispersion as the honeycomb lattice. Thus, the bottom of lowest-energy band locates at $\Gamma$.

We now combine the Kagome lattice with the honeycomb lattice to form the HK lattice. The HK lattice has a spatial inversion symmetry with a $D_{6h}$ point group symmetry. Since the s orbital and $p_z$ orbital is symmetric or anti-symmetric with respect to the *xy*-mirror plane, there is no interaction between s and $p_z$ orbitals. Thus, the band structure of the HK lattice will be simply the superimposition of the band structures of the Kagome and honeycomb lattices. If the bottommost band of the Kagome lattice has a lower energy than the topmost band of the honeycomb lattice at $\Gamma$, there will be band inversion between the s and $p_z$ bands in the HK lattice. Note that these two states have the even parity, suggesting that the band inversion has a different meaning from the usual band inversion that happens between bands with opposite parities. Since the s band and $p_z$ band are even or odd with respect to the *xy* mirror plane, the band crossing points will form a closed node-line ring. Thus, the 2D NLS state protected by the mirror symmetry may be realized in the HK lattice.

**Realization of 2D NLS state in real materials.** In the following, we will design 2D NLS material based on the HK $A_3B_2$ lattice. In order to obtain a 2D NLS state in a HK lattice, the valence band and conduction band should be contributed by $p_z$ and s states, respectively. In usual compounds, the occupied p state and empty s state originates from the anion and cation, respectively. Therefore, the honeycomb B-sites should be anions, while the kagome A-sites should be cations. Since -6 valence state of an anion is rare, the valence states of A-cation and B-anion of a stable $A_3B_2$ compound should be +2 and -3, respectively. After extensive tests, we find that if the A-cation is a group-IIB element (e.g., Zn, Cd and Hg) and B-anion is a group VA element (e.g., N, P, As, Sb and Bi), the $A_3B_2$ compound might be a 2D NLS. With first principle calculations, we obtain the relaxed structures and find that 2D NLS state can be present in five $A_3B_2$ compounds (i.e., $Zn_3Bi_2$, $Cd_3Bi_2$, $Hg_3As_2$, $Hg_3Sb_2$ and $Hg_3Bi_2$). In addition, although planar $Cd_3As_2$ with the relaxed structure is a normal semiconductor, it becomes a 2D NLS when 1% tensile strain is applied. As can be seen from the band structures of these compounds [Fig. 2(a) and supplementary Fig.

S1], there are two bands crossing each other near Γ point at the Fermi level. The fat band representation clearly shows that the conduction band is contributed by s orbital of cation and the valence band is contributed by $p_z$ orbital of anion. It is found that the degree of the band inversion increases when cation and anion becomes heavier. For example, the inversed energies of the Hg system are 0.841, 0.982 and 1.127 eV for $Hg_3As_2$, $Hg_3Sb_2$ and $Hg_3Bi_2$, respectively. This is because the s level decreases with atomic number of the cation element, while the p level increases with atomic number of the anion element. Since the local density approximation usually underestimates the band gap, we adopt a more reliable hybrid functional HSE06 to confirm that $Hg_3As_2$ is still a 2D NLS (see supplementary Fig. S2). Interestingly, we find that the optical absorption of HK $Hg_3As_2$ below 2 eV is very weak (see supplementary Fig. S3). This is because both VBM and CBM states have even parity, thus the direct optical transition between them is forbidden. Due to the linear band dispersion near the node-line ring, the conductivity in HK $Hg_3As_2$ is expected to be rather high. This suggests that 2D NLS can be a promising candidate for transparent conductors in touch screens and solar cells [35, 36].

So far, the effect of SOC is neglected. After including the SOC in the density functional calculations, a small band gap will be opened around the node-line ring (supplementary Table S1). We can understand this from the symmetry analysis. One only needs to check whether the gap will be opened along the two high-symmetry lines starting from Γ (i.e., $\Gamma \to K$ and $\Gamma \to M$). For the k-points on the two lines (except for the end points), the symmetry group is the double group of $C_{2v}$. Since the double group of $C_{2v}$ only has a 2D irreducible representation $\Gamma_5$, suggesting that both CBM and VBM states much belong to the same irreducible representation $\Gamma_5$ and they could interact with each other, resulting in a band gap opening. For $Hg_3As_2$, the band gap is 34 meV [see Fig. 2b]. This value is smaller than the SOC-induced gap opening (about 62 meV) in a 3D NLS system $Cu_3NPd$ [24]. Note that the SOC-induced band gap opening can be tuned by changing the elements. For example, the SOC-induced band gap in $Cd_3As_2$ with 1% tensile strain is only 0.2 meV.

To gain more insight into the origin of the 2D NLS and SOC-induced gap opening in $A_3B_2$ compounds, we construct the effective $k \cdot p$ Hamiltonian with the invariant method [37]. To be more specific, we consider $Hg_3As_2$ as a typical example. According to the LDA calculation (without SOC), the VBM and CBM states at $\Gamma$ are mainly contributed by Hg $|s\rangle$ and As $|p_z\rangle$ orbitals, respectively. The VBM state is a bonding state between the three Hg s orbitals, while the CBM state is an anti-bonding state between the As $p_z$ orbitals:

$$|S^+\rangle = \frac{1}{\sqrt{3}}(|Hg_1, s\rangle + |Hg_2, s\rangle + |Hg_3, s\rangle),$$

$$|P_z^+\rangle = \frac{1}{\sqrt{2}}(|As_1, p_z\rangle - |As_2, p_z\rangle),$$

where the "+" sign indicating that both states have even parity. The irreducible representations of $|S^+\rangle$ and $|P_z^+\rangle$ are $\Gamma_1^+$ and $\Gamma_3^+$ of $D_{6h}$, respectively. We will use $|S^+\rangle$ and $|P_z^+\rangle$ as the basis to construct low-energy effective Hamiltonian around the $\Gamma$ point. The effective Hamiltonian obtained from the invariant method reads:

$$H(k) = \begin{bmatrix} \varepsilon(k) + M(k) & 0 \\ 0 & \varepsilon(k) - M(k) \end{bmatrix}$$

where $\varepsilon(k) = C_0 + C_1(k_x^2 + k_y^2)$ and $M(k) = M_0 + M_1(k_x^2 + k_y^2)$. It is clear that these two states are decoupled with the eigenvalues $E(k) = C_0 \pm M_0 + (C_1 \pm M_1)(k_x^2 + k_y^2)$. It can be seen that if $M_0 M_1 > 0$, the system is a normal insulator. Otherwise, a band inversion can happen. In this case, the two bands will cross each other at the k-points which satisfy $k_x^2 + k_y^2 = -M_0/M_1$. This means that the crossing points form a node-line ring [see Fig. 2c], in agreement with the first-principles result.

In the case of SOC, we now have four basis functions, namely, $|S^+ \uparrow\rangle, |P_z^+ \uparrow\rangle, |S^+ \downarrow\rangle, |P_z^+ \downarrow\rangle$. $|S^+ \uparrow, \downarrow\rangle$ and $|P_z^+ \uparrow, \downarrow\rangle$ belong to 2D $\Gamma_7^+$ and $\Gamma_8^+$ irreducible representations of double group of $D_{6h}$, respectively. Here, besides the

point group symmetry, time reversal symmetry is also taken into account to derive the effective Hamiltonian up to the third order of **k** around Γ:

$$H(k) = \begin{bmatrix} \varepsilon(k)+M(k) & 0 & 0 & B(k) \\ 0 & \varepsilon(k)-M(k) & -B(k) & 0 \\ 0 & -B(k)^* & \varepsilon(k)+M(k) & 0 \\ B(k)^* & 0 & 0 & \varepsilon(k)-M(k) \end{bmatrix},$$

where $B(k) = A_0\left[i\left(k_x^2 - k_y^2\right) - k_x k_y\right]$ ($A_0$ characterizes the magnitude of the SOC effect). It is interesting to see that the effective Hamiltonian does not contain the linear term and third order term of **k** because the basis functions have the same parity. This is different from the case of usual topological systems in which there are linear **k** terms [37]. Due to the presence of both time-reversal and inversion symmetry, there are two double degenerate bands with eigenvalues $E(k) = \varepsilon(k) \pm \sqrt{M(k)^2 + |B(k)|^2}$. Interestingly, there is no four-fold degenerate point in the whole Brillouin zone since it is impossible to satisfy $M(k)=0$ and $B(k)=0$ simultaneously, i.e., a band gap will be opened at the node-line ring by the SOC effect [see Fig. 2d]. The absence of the linear **k** terms in our Hamiltonian naturally explains the small band gaps induced by SOC. Our $k \cdot p$ analysis thus explains the SOC-induced gap opening in the LDA+SOC band structure.

We find that the insulating state of $Hg_3As_2$ with the HK lattice induced by the SOC is a 2D TCI state [38, 39]. Qualitatively, this can be understood by examining the band structure without considering SOC. From the LDA band structure and Table S1, band inversion only takes place at the Γ point instead of on the other three time-reversal points. Note that the meaning of "band inversion" here is different from that in the field of topological insulators: In the latter case, "band inversion" usually means the inversion between the eigenstates with opposite parity; while in our context, both eigenstates (s and $p_z$ states) have the same even parity but opposite eigenvalues respecting to the *xy*-mirror plane. Since the band inversion relating to the mirror symmetry occurs only once, it is expected that the system will become a 2D TCI state after a band gap is opened by SOC. Interestingly, the band inversion mechanism in our case is different from the case in monolayer IV-VI (e.g., PbSe) where the TCI behavior is due to the band inversion with respect to both spatial inversion and *xy*-mirror plane at X and Y points [28, 40, 41]. We support our above argument by numerically computing the mirror Chern number with a Slater-Koster TB model. In this TB model, we include one s orbital for a Hg ion and three p orbitals for an As ion (not counting spin). The on-site SOC on As p orbitals is included to open the band gap. We find that $Hg_3As_2$ has a non-zero mirror Chern number of 2, suggesting that it is a TCI after SOC is considered. We further confirm this result by computing the edge states of the semi-infinite $Hg_3As_2$ nanoribbons with both armchair and zigzag terminations. The recursive method is adopted to compute the surface local density of states [42]. Since the presence of the edges does not break the mirror symmetry with respect to the *xy*-plane, it is expected that there will exist nontrivial helical edge states as a characteristic feature of TCI [43]. This is indeed the case. As can be seen from Fig. 3 and Fig. S4, there are two helical edge states within the bulk band gap for different terminations. The number of the edge states is in agreement with the calculated mirror Chern number.

We now check the structural stability of the $A_3B_2$ compounds with the HK lattice. By performing global structure optimization [44], we find that the buckled $Hg_3As_2$ structure [as shown in Fig. 4a] with the HK lattice has the lowest energy among all structures with the thickness less than 0.8 Å. The lateral lattice constant of buckled $Hg_3As_2$ is 8.502 Å, and Hg and As atoms locate in two planes with a distance of 0.732 Å. The computed phonon frequencies [45, 46] (as shown in supplementary Fig. S5) indicate that buckled HK $Hg_3As_2$ is dynamically stable. The LDA band structure shown in Fig. 4c indicates that buckled HK $Hg_3As_2$ is a semiconductor with a LDA direct gap of 1.629 eV. The conduction and valence bands are contributed by Hg-s orbital and As-$p_z$ orbitals, respectively. Therefore, buckled HK $Hg_3As_2$ is a normal insulator instead of a 2D NLS.

It is found that the band gap of buckled HK $Hg_3As_2$ decreases quickly as the thickness goes thinner [see Fig. 4b]. When the thickness is less than 0.3 Å, the gap closes. From the LDA band structure of HK $Hg_3As_2$ with a 0.3 Å buckling [see Fig.

4d], we find that the conduction and valence bands cross each other along six special paths of the momentum space, i.e., $\Gamma \to K$ and $\Gamma \to K'$, but a tiny gap (about 5 meV) is opened along other directions (e.g., $\Gamma \to M$). To further understand this, we analyze the wave function symmetries along the $\Gamma \to K$ and $\Gamma \to M$. Along the $\Gamma \to K$, the irreducible representations of CBM and VBM are $A'$ and $A''$ of the Cs group, respectively. While for the $\Gamma \to M$, they belong to the same irreducible representation $A'$ of the Cs group, resulting in a gap opening. This can be seen more clearly from the 3D band structure shown in supplementary Fig. S6. Our result suggests that a small buckling will transform the 2D NLS state of the HK lattice into a 2D DSM state. An effective $k \cdot p$ Hamiltonian is derived to understand the effect of buckling on the electronic structure (see supplementary material). We find an additional third-order term of k due to the buckling, which explains why the band gap is opened along $\Gamma \to M$, but the gap is tiny. Since the band opening by the buckling is small, HK $Hg_3As_2$ with a small buckling can still be seen as a 2D NLS at room temperature.

We propose that the thickness of buckled HK $Hg_3As_2$ can be tuned by applying pressure on $Hg_3As_2$ sandwiched by two insulating BN monolayers. The pressure (about 9 MPa) needed for turning the normal insulating state into the 2D NLS state can be easily achieved experimentally. The band structure of the three-layer system under an external pressure of 11 MPa is shown in Fig. S7. We can see that the 2D NLS remain intact despite of the presence of BN layers since the states near the Fermi level is mainly contributed by Hg-s and As-$p_z$ orbital.

In summary, in this work we propose for the first time the concept of the 2D NLS. A newly constructed HK lattice is predicted to display the 2D NLS behavior due to the band inversion with respect to the mirror symmetry. This 2D NLS state can be realized in the real $A_3B_2$ (A is a group-IIB cation and B is a group-VA anion) compounds (such as $Hg_3As_2$) with the HK lattice. Since the band inversion occurs between two bands with the same parity, this peculiar 2D NLS could be used as transparent conductors. When SOC is included, HK $Hg_3As_2$ becomes a small-gap topological crystal insulator. The effect of buckling on the electronic properties of HK $Hg_3As_2$ is further discussed. Since peculiar quantum Hall effect has been observed in graphene with 2D Dirac band dispersion [47], it is expected that the 2D NLS state might lead to novel quantum Hall effect and a new route to achieve 2D high-temperature superconductivity. Our work also suggests a new route without involving states with opposite parity to design topological materials.

Work at Fudan was supported by NSFC (11374056), the Special Funds for Major State Basic Research (2012CB921400, 2015CB921700), Program for Professor of Special Appointment (Eastern Scholar), Qing Nian Bo Jian Program, and Fok Ying Tung Education Foundation. J.L. L. and W. L. contributed equally to this work.

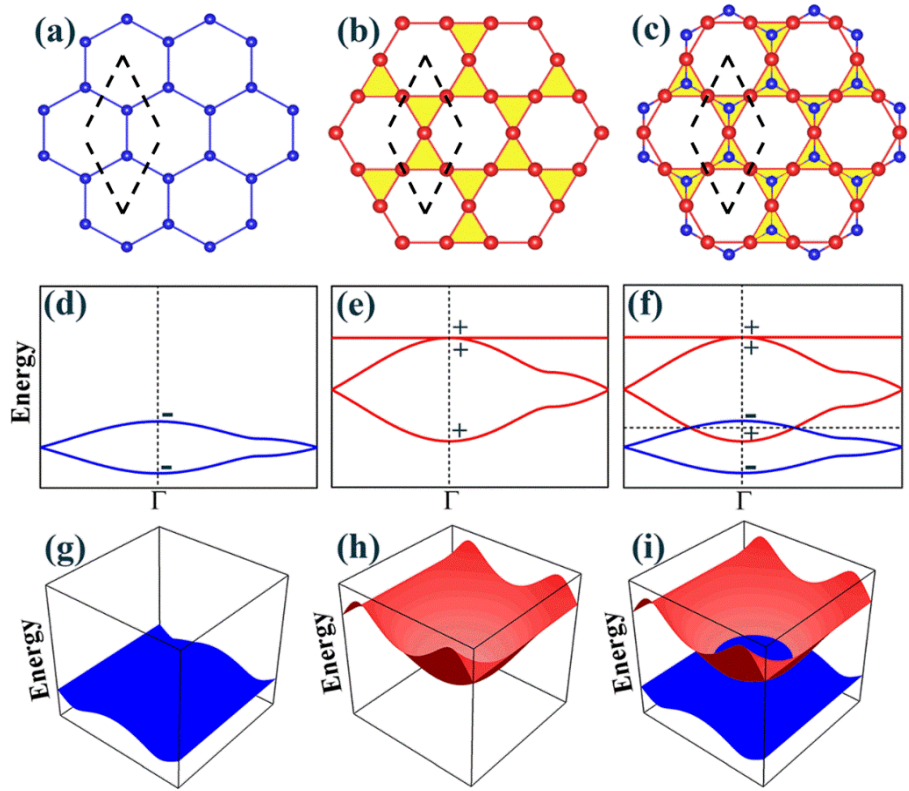

**FIG. 1** 2D node-line semimetal in the honeycomb-kagome lattice. (a-c) The geometrical structures of the honeycomb, kagome, and honeycomb-kagome (HK) lattices. (d-f) The 2D band structures of honeycomb, kagome, and HK lattices. (g-i) The corresponding 3D band structures. The dashed lines show the unit cell. The *x*-axis is along the direction of a lateral lattice vector. There is a $p_z$ (s) orbital on each honeycomb (kagome) lattice point. The band structures are computed with $E_p = -4.6$, $t_p = 0.3$ for the honeycomb lattice and $E_s = -1.8$, $t_s = 0.6$ for the kagome lattice.

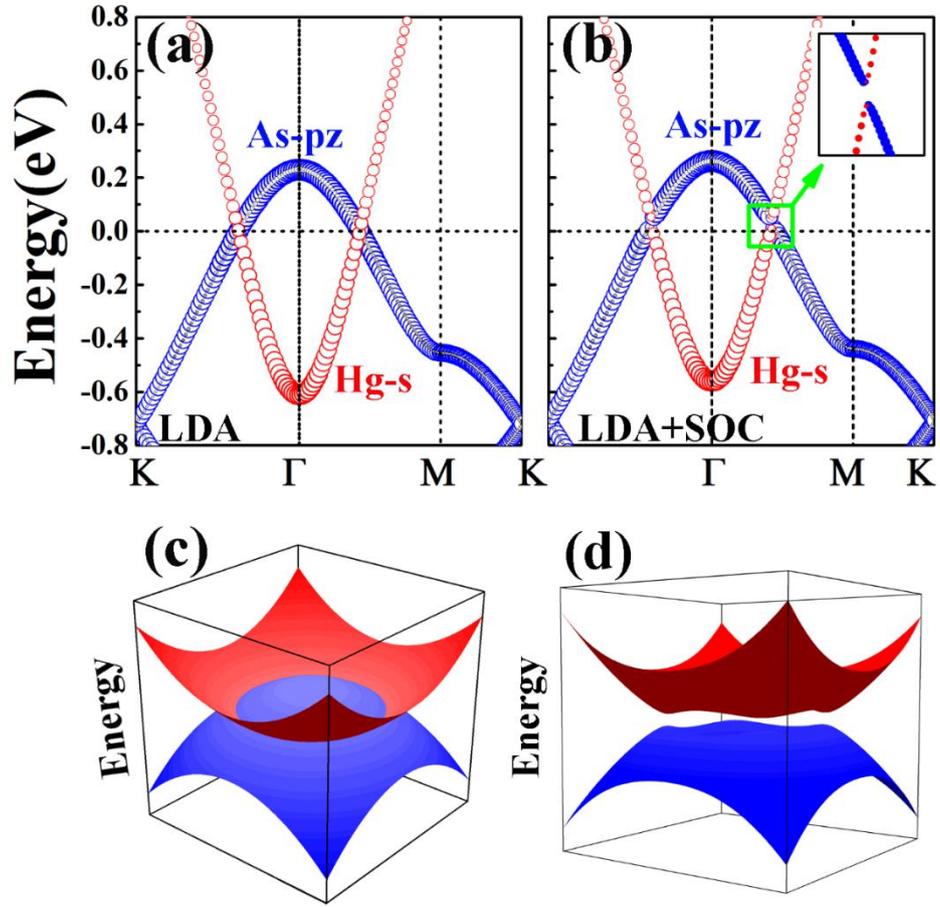

**FIG. 2** Band structures of planar Hg$_3$As$_2$ with the HK lattice. (a-b) The 2D band structures calculated by LDA and LDA+SOC, respectively. (c-d) The corresponding 3D band structures from the $k \cdot p$ theory. In (c), $C_0 = -1, C_1 = 1, M_0 = -1, M_1 = 1$ is employed. Besides these parameters, a large value (0.5) for $A_0$ is adopted in (d) for clarity.

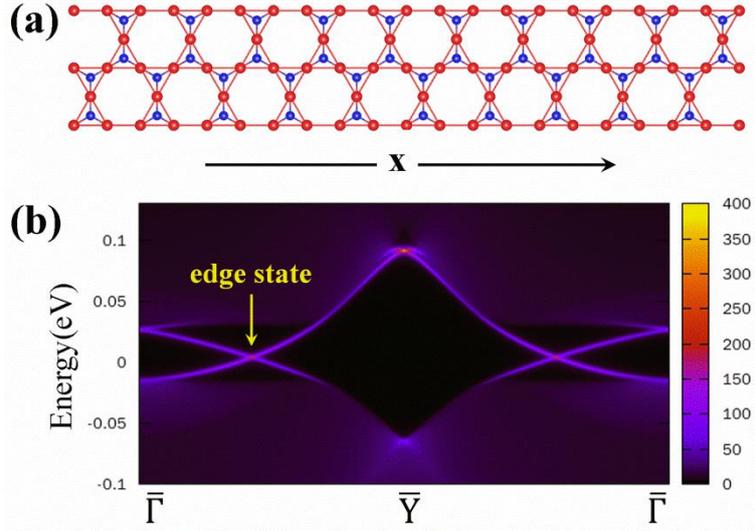

**FIG. 3** The edge states of armchair ribbon of $Hg_3As_2$ with the HK lattice. (a) The geometrical structures. (b) Energy and momentum-dependent local density of states of the semi-infinite armchair ribbon computed through the Green function method. The SOC effect has been included. Two edge states in the bulk band gap of the 2D TCI can be clearly seen.

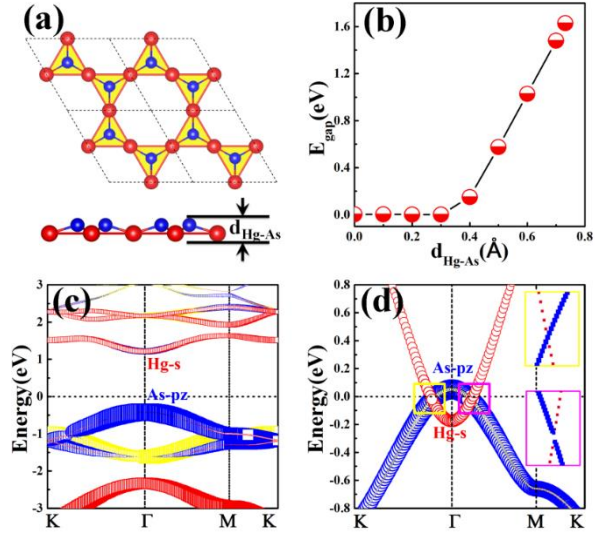

**FIG 4** Geometrical and electronic structures of buckled HK $Hg_3As_2$. (a) Top and side view of the structure of buckled $Hg_3As_2$. (b) Band gaps of buckled $Hg_3As_2$ as a function of thickness from the LDA calculations. (c) Band structure of optimized buckled $Hg_3As_2$. (d) Band structure of $Hg_3As_2$ with a thickness of 0.3 Å. Insets: top panel for the magnified view of the bands along the line K→Γ, bottom panel for the case of Γ→M.